# Microwave assisted resonant domain wall nucleation in permalloy nanowires

Masamitsu Hayashi[*], Yukiko K. Takahashi and Seiji Mitani

*National Institute for Materials Science, Tsukuba 305-0047, Japan*

We have designed a system to study microwave assisted domain wall nucleation in permalloy nanowires. We find a substantial decrease in the nucleation field when microwave fields are applied, in comparison to pulse fields. A clear resonance peak is observed in the frequency dependence of the nucleation field, which coincides with the uniform mode ferromagnetic resonance frequency. Owing to the well-defined nucleation process, the switching field distribution is small in contrast to previous reports. Our results show that localized microwave field provides an efficient tool for injecting domain walls into magnetic nanowires.

[*]E-mail address: hayashi.masamitsu@nims.go.jp

Recent progress in the understanding of spin transfer torque and its related phenomena enabled development of domain wall based memory[1] and logic devices[2]. In particular, a magnetic shift register mimicking the so called "Racetrack memory"[1] has been demonstrated using localized magnetic fields to inject domain walls and spin polarized current to move them[3-5]. One challenge of exploring materials for domain wall based device applications is the difficulty in injecting domain walls into the nanowire since it typically requires large localized magnetic field[3,6]. Solid understanding of the nucleation process is thus required for further development of such devices.

Microwave assisted magnetization reversal (MAMR) provides a unique tool to lower the magnetization switching field[7,8]. The concept has been demonstrated in a number of systems[7,9-18] but the understanding of the underlying physics is still developing. In most of the systems, it is likely that the magnetization reversal process involves nucleation and propagation of domain walls (exception is the coherent rotation of magnetic moments[12]). Both domain wall (reversed domain) nucleation and propagation can be enhanced by the application of microwave field. However, its response may differ depending on the applied microwave frequency and thus may cause broadening of the switching field[9,10,17,18]. Thus to gain further insight into MAMR, it is essential to distinguish the effect of microwave field application on domain wall nucleation[19] and domain wall propagation[20,21].

Here we design a system to study the effect of microwave field application on a well-defined domain wall nucleation process in permalloy naonwires. Substantial drop in the nucleation field is observed for a microwave field frequency of ~3 GHz. We find a clear resonance peak in the frequency dependence of the nucleation field, in which the peak position corresponds to the ferromagnetic resonance uniform excitation mode. The domain wall nucleation field is reduced by ~50% at the resonance condition compared to the *pulse* field excitation, which shows the high efficiency of microwave wave assisted domain wall nucleation process.

Permalloy ($Ni_{81}Fe_{19}$) nanowires are made from film stack of 10 $Ni_{81}Fe_{19}$|5 Ru (units in nm) deposited on single crystal MgO substrate using magnetron sputtering. Figures 1(a) shows an optical microscopy image of the device and schematic illustration of the measurement setup. Electron beam (e-beam) lithography and Ar ion milling are used to define the magnetic nanowire. Typical width and length of the nanowire are 600 nm and 32 μm, respectively. Contacts to the nanowire are made from 5 Ta|85 Au using e-beam lithography and a liftoff process. An isolation pad, 30 nm thick $SiO_2$ layer, is inserted between the permalloy nanowire and the field generation lines (lines A and B in Fig. 1(a)). The width of line A parallel to the permalloy nanowire is set to ~2 μm, large enough to cover the 600 nm wide nanowire. A signal generator or a pulse generator is connected to field generation lines A and/or B to apply localized magnetic field. Note that the Oersted field generated at line A (B), when current is passed along the line, is pointing transverse (parallel) to the wire's long axis. The rise time of the pulse generator is typically ~500 ps. We use a 7 GHz bandwidth digitizing oscilloscope, connected to line C, D in Fig. 1(a), to monitor the change in the voltage drop across this section (C-D) of the wire. DC current (~500 μA) is supplied to the nanowire via a bias tee in order to read out the wire's resistance through the scope voltage.

Propagation of domain walls are studied using time resolved voltage measurements. We cycle the external magnetic field to first reset the magnetic state of the wire by applying a large magnetic field -$H_{SAT}$, and then reduce the field to H in order to nucleate and drive a domain wall along the wire, and finally apply -$H_{SAT}$ to avoid the electromagnet to get demagnetized. Positive field points along the +x direction in Fig. 1(a). When the field is at H, we apply a voltage pulse (180 ns long) from the pulse generator to apply a localized Oersted field around the field generation line. The direction of the Oersted field, determined by the polarity of the voltage pulse, is set to point along H (for pulses applied to line B) to assist the magnetization reversal process. We monitor the scope voltage when the voltage pulse is applied. The field cycling is repeated for 256-512 times in order to increase the signal to noise ratio.

We first examine the domain wall nucleation field when the local Oersted field is directed parallel to the wire's long axis. Here, a pulse generator is connected to line B. Figures 1(b) and 1(c) show representative scope traces captured when the amplitude of the voltage pulse ($V_P$) applied to the field generation line is varied (pulse length is fixed to 180 ns). Due to the anisotropic magnetoresistance (AMR) of permalloy, the scope signal drops from zero to a negative value when a domain wall enters section C-D[22]. For $V_P$=-0.8 V (Fig. 1(b)), we observe signatures of domain wall nucleation and injection into section C-D for all H shown. However, for $V_P$=-0.5 V (Fig. 1(c)), no domain wall injection is observed when H is 35 Oe. Note that the propagation field of this wire, i.e. the minimum field needed to move a domain wall along the wire, is ~ 5 Oe. These results indicate that when H=35 Oe and $V_P$=-0.5 V, combination of H and the Oersted field is not enough to nucleate a domain wall, thereby resulting in no domain wall passing by section C-D.

We define the "assisting field" as the minimum H needed to assist the domain wall nucleation process. The assisting field is plotted in Fig. 1(d) as a function of the current passed along the line B. For smaller voltage pulse, larger assisting field (H) is needed to nucleate a domain wall. As a reference, the dotted line in Fig. 1(d) represent the field (H) needed to reverse the wire magnetization with quasi-static field. This field (defined as $H_S$~77 Oe hereafter) represents the minimum field needed to nucleate a reversed domain (i.e. a domain wall) likely from the edges of the wire. Once a domain wall is created, the wall can propagate along the wire as long as the field is larger than ~5 Oe, which results in magnetization reversal. Note that the total field required to locally nucleate a domain wall, i.e. the nucleation field, is larger than $H_S$ in all cases. For example, the nucleation field for $V_P$=-0.8 V is ~125 Oe, which includes the assisting field of ~15 Oe and the current induced Oersted field at line B of ~110 Oe (the Oersted field is calculated using the Biot-Savart law[23]).

We next turn to the effect of the transverse field (perpendicular to the wire's long axis and H) on the domain wall nucleation process. Here we use line A to generate the transverse Oersted field (pulse generator is connected to line A). Similar to the method described in Fig. 1(d), we extract the assisting field H needed to nucleate a domain wall. The assisting field is represented by the red circles in Fig. 2(e), plotted as a function of current passed along line A. As observed with the parallel field, the assisting field decreases with the magnitude of current.

These assisting fields are compared with those obtained using the microwave Oersted field. A signal generator (microwave source) is connected to line A and a pulse generator is connected to line B. Since the rise time of the microwave source is slow (more than ~1 μs), the source is turned on throughout the field cycling. The microwave Oersted field is small enough such that it does not affect the saturation process when $H_{SAT}$ is applied. A voltage pulse (-2.0 V, 180 ns) is applied from line B to nucleate a domain wall, as in Fig. 1(d). Exemplary scope traces are shown in Fig. 2(a-c) for three different microwave frequencies $f_{RF}$. The microwave power $P_{RF}$ is fixed to 0 dBm (Vrms~0.22 V). For microwave frequencies of 0.1 GHz (Fig. 2(a)) and 6.0 GHz (Fig. 2(c)), signature of domain wall nucleation and propagation is observed for all H shown.   In contrast, when $f_{RF}$=3.0 GHz (Fig. 2(b)), drop in the scope signal associated with the presence of domain wall in section C-D is absent when H is above 35 Oe.   This indicates that the microwave field creates a domain wall and reverses the magnetization direction of the wire via domain wall propagation *prior* to the injection of the voltage pulse to line B when the field is at H (nucleation of the wall by the microwave field is confirmed by monitoring the microwave field induced nucleation process in real-time (results not shown)). Note that the propagating spin waves generated at line A via parametric pumping can influence domain wall propagation[24-27]   However, using a wire with different configuration of the field generation lines (i.e. line A is placed to the right of line C-D) we observe no change in the nucleation field and thus consider that the effect of spin waves on the nucleation process is negligible.

The dependence of the amplitude of the scope signal, as defined in Fig. 2(a), is plotted as a function of H for various $f_{RF}$ in Fig. 2(d).   Since each scope signal is an average of a few hundred independent measurements, the amplitude represents the probability of a domain wall moving into section C-D.   In addition, the amplitude also reflects the mean size of the domain wall as it propagates.   The gradual increase in the amplitude with the field reflects the variation of the wall width[28-30]: the wall structure typically gets elongated at higher fields[31].   The Walker breakdown field here is close to the domain wall propagation field (~5 Oe).   For a given $f_{RF}$, we look for H above which the amplitude drops to zero.   This field is defined again as the "assisting field" and is plotted in Fig. 2(e) for $f_{RF}$=0.1 GHz and 3.0 GHz as a function of the current applied to line A.   The current shown here is the amplitude of the sinusoidal current (not the rms value).   Interestingly, when $f_{RF}$=3.0 GHz, the assisting field needed is, in most cases, smaller than that when a pulse Oersted field is used. This trend is pronounced when the current applied to the field generation line is smaller. For $f_{RF}$=0.1 GHz, the required assisting field is always larger than that of the pulse excitation (the difference between the two is rather large and to understand the reason behind this requires further investigation).

Since the width of the field generation lines are different between lines A and B, the Oersted field generated from each line at a given current is different. We thus calculate the Oersted field in each geometry and attempt to directly compare the efficiency of domain wall nucleation between the two. These results are summarized in Fig. 3(a). Here the "nucleation field", defined by the sum of the assisting field and the calculated Oersted field, is plotted as a function of the Oersted field. Nucleation fields for the pulse excitation using lines A (Oersted field transverse to the easy axis) and B (Oersted field is parallel to the easy axis) are shown by the red circles and black squares, respectively. The triangles in Fig. 3(a) represent the nucleation fields for the microwave excitation from line A (transverse Oersted field). The nucleation field is always smaller when the Oersted field is pointing transverse to the easy axis as compared to the parallel Oersted field. The nucleation field is the smallest among all the geometries used when microwave current of $f_{RF}$=3.0 GHz is applied to line A (transverse Oersted field). Note that the nucleation field tends to increase when the Oersted field amplitude is increased, particularly for the parallel pulse excitation (squares). This feature may depend on the details of the nucleation process and further investigation is required to clarify its origin.

To further illustrate the effect of the microwave field on the nucleation field, we show the excitation frequency dependence of the nucleation field in Fig. 3(b). The microwave power is set to 0 dBm (Oersted field of ~27 Oe), as shown by the vertical dashed line in Fig. 3(a). We clearly observe a drop in the nucleation field when $f_{RF}$~3.0 GHz, indicating a resonance effect on the nucleation process. This is in contrast to previous results where the *nucleation field* shows little dependence with the microwave field application[19]. Note that the change in the nucleation field with respect to $f_{RF}$ is asymmetric: such effect has been predicted by micromagnetic simulations but not been observed experimentally[19].

To understand the origin of the "resonance frequency" at which the nucleation field takes its minimum, we measure the ferromagnetic resonance (FMR) in a similar nanowire (same wire width). A coplanar waveguide (signal line width: 2 μm, ground line width: 4 μm, distance between the two lines: 1.4 μm) made of 5 Ta|85 Au (units in nm) is formed on top of a permalloy Hall bar (wire width: 600 nm, wire length: 60 μm, Hall bar width: 500 nm). See Fig. 3(c) for the experimental setup. A 30 nm thick $SiO_2$ layer is inserted between the permalloy Hall bar and the waveguide to avoid shorting the Hall bar with the waveguide. Direct current (100 uA) is applied to the Hall bar and we use the AMR effect to probe the dynamics of the magnetic moments[32]. Sinusoidal current is applied to the waveguide to generate in-plane oscillating field to the permalloy layer. The inset of Fig. 3(d) shows an exemplary in-plane field dependence of the permalloy Hall bar resistance. Here, a microwave current (9.5 GHz, 6 dBm) is applied to the waveguide. Large negative peaks appearing at ~±750 Oe is the uniform excitation mode and the smaller peaks observed at smaller fields are likely to do with the quantized spin wave modes[33]. The peak fields of the uniform mode (absolute mean of the positive and negative peak fields) are plotted as a function of the sinusoidal current frequency in Fig. 3(d). At small magnetic fields (~27 Oe in Fig. 3(a)), the frequency of the uniform mode is ~3 GHz, similar to what we find for the "resonance frequency" shown in Fig. 3(b).

A number of studies have reported that microwave assisted magnetization switching is rather stochastic[9] and the switching field shows broad distribution[9,10,17,18]. This is likely to do with the presence of multiple nucleation sites with different internal fields (and thus different resonant frequencies) that can contribute to the switching process. In addition, the influence of microwave field application on domain wall propagation may also add to the switching field distribution[20,21]. As evident from the plots in Fig. 2(d), the amplitude (which is proportional to one minus the nucleation probability) shows an abrupt drop above a certain H for each $f_{RF}$ which indicates that the switching field distribution is rather small. Since we use a localized magnetic field, we consider that domain wall nucleation always occurs from a fixed position, likely at the edge of the field generation lines, which contributes to the small distribution in the nucleation field. In addition, the effect of microwave field application on domain wall propagation is excluded in our system, which eliminates another factor that adds to the broadening of the switching field.

In summary, we have studied domain wall nucleation in permalloy nanowires under pulse and microwave Oersted field excitations. Contrary to previous reports where microwave field likely acted on both domain wall nucleation and propagation, here we design a system to study only the former process. We find that the nucleation field is reduced by a factor of ~2 when a resonant transverse microwave Oersted field, whose amplitude is roughly a quarter of the off-resonant nucleation field, is applied. The resonance frequency is ~3 GHz, which is close to the ferromagnetic resonance frequency of the uniform excitation mode. Owing to the well-defined nucleation process, the nucleation field upon microwave field application shows small distribution. Our results show that localized microwave fields are effective in injecting domain walls in magnetic nanowires.

**Acknowledgments**

This work is partly supported by a Grant-in-Aid for Scientific Research (No. 22760015) from MEXT, Japan.

**References**

[1] S. S. P. Parkin, M. Hayashi, and L. Thomas, Science **320**, 190 (2008).

[2] D. A. Allwood, G. Xiong, C. C. Faulkner, D. Atkinson, D. Petit, and R. P. Cowburn, Science **309**, 1688 (2005).

[3] M. Hayashi, L. Thomas, R. Moriya, C. Rettner, and S. S. P. Parkin, Science **320**, 209 (2008).

[4] K.-J. Kim, J.-C. Lee, S.-J. Yun, G.-H. Gim, K.-S. Lee, S.-B. Choe, and K.-H. Shin, Appl. Phys. Express **3**, 083001 (2010).

[5] D. Chiba, et al., Appl. Phys. Express **3**, 073004 (2010).

[6] F. U. Stein, L. Bocklage, T. Matsuyama, and G. Meier, Appl. Phys. Lett. **100**, 192403 (2012).

[7] C. Thirion, W. Wernsdorfer, and D. Mailly, Nat. Mater. **2**, 524 (2003).

[8] J. G. Zhu, X. C. Zhu, and Y. H. Tang, IEEE Trans. Magn. **44**, 125 (2008).

[9] Y. Nozaki, K. Tateishi, S. Taharazako, M. Ohta, S. Yoshimura, and K. Matsuyama, Appl. Phys. Lett. **91**, 122505 (2007).

[10] T. Moriyama, R. Cao, J. Q. Xiao, J. Lu, X. R. Wang, Q. Wen, and H. W. Zhang, Appl. Phys. Lett. **90**, 152503 (2007).

[11] P. M. Pimentel, B. Leven, B. Hillebrands, and H. Grimm, J. Appl. Phys. **102**, 063913 (2007).

[12] G. Woltersdorf and C. H. Back, Phys. Rev. Lett. **99**, 227207 (2007).

[13] J. Podbielski, D. Heitmann, and D. Grundler, Phys. Rev. Lett. **99**, 207202 (2007).

[14] X. L. Fan, Y. S. Gui, A. Wirthmann, G. Williams, D. S. Xue, and C. M. Hu, Appl. Phys. Lett. **95**, 062511 (2009).

[15] C. Nistor, K. Sun, Z. H. Wang, M. Z. Wu, C. Mathieu, and M. Hadley, Appl. Phys. Lett. **95**, 012504 (2009).

[16] J. Topp, D. Heitmann, and D. Grundler, Phys. Rev. B **80**, 174421 (2009).

[17] T. Yoshioka, T. Nozaki, T. Seki, M. Shiraishi, T. Shinjo, Y. Suzuki, and Y. Uehara, Appl. Phys. Express **3**, 013002 (2010).

[18] S. Okamoto, N. Kikuchi, O. Kitakami, T. Shimatsu, and H. Aoi, J. Appl. Phys. **109**, 07B748 (2011).

[19] H. T. Nembach, H. Bauer, J. M. Shaw, M. L. Schneider, and T. J. Silva, Appl. Phys. Lett. **95**, 062506 (2009).

[20] E. Schlomann, IEEE Trans. Magn. **11**, 1051 (1975).

[21] A. Krasyuk, F. Wegelin, S. A. Nepijko, H. J. Elmers, G. Schonhense, M. Bolte, and C. M. Schneider, Phys. Rev. Lett. **95**, 207201 (2005).

[22] M. Hayashi, L. Thomas, Y. B. Bazaliy, C. Rettner, R. Moriya, X. Jiang, and S. S. P. Parkin, Phys. Rev. Lett. **96**, 197207 (2006).

[23] T. J. Silva, C. S. Lee, T. M. Crawford, and C. T. Rogers, J. Appl. Phys. **85**, 7849 (1999).

[24] S. M. Seo, K. J. Lee, H. Yang, and T. Ono, Phys. Rev. Lett. **102**, 147202 (2009).

[25] D.-S. Han, S.-K. Kim, J.-Y. Lee, S. J. Hermsdoerfer, H. Schultheiss, B. Leven, and B. Hillebrands, Appl. Phys. Lett. **94**, 112502 (2009).

[26] M. Jamali, K. J. Lee, and H. Yang, New J. Phys. **14**, 033010 (2012).

[27] J. S. Kim, M. Stark, M. Klaui, J. Yoon, C. Y. You, L. Lopez-Diaz, and E. Martinez, Phys. Rev. B **85**, 174428 (2012).

[28] G. S. D. Beach, C. Nistor, C. Knutson, M. Tsoi, and J. L. Erskine, Nat. Mater. **4**, 741 (2005).

[29] M. Hayashi, L. Thomas, C. Rettner, R. Moriya, and S. S. P. Parkin, Nat. Phys. **3**, 21 (2007).

[30] J.-Y. Lee, K.-S. Lee, S. Choi, K. Y. Guslienko, and S.-K. Kim, Phys. Rev. B **76**, 184408 (2007).

[31] M. Hayashi, S. Kasai, and S. Mitani, Appl. Phys. Express **3**, 113004 (2010).

[32] N. Mecking, Y. S. Gui, and C. M. Hu, Phys. Rev. B **76**, 224430 (2007).

[33] J. Jorzick, S. O. Demokritov, B. Hillebrands, M. Bailleul, C. Fermon, K. Y. Guslienko, A. N. Slavin, D. V. Berkov, and N. L. Gorn, Phys. Rev. Lett. **88**, 047204 (2002).

**Figure captions**

Fig. 1. (a) Optical microscopy image of the nanowire device along with schematic illustration of the experimental setup. The area trimmed by the dotted line indicates where the 30 nm thick $SiO_2$ isolation pad is inserted. (b,c) Typical waveforms captured by the oscilloscope for various fields H and pulse amplitudes: (b) -0.8 V and (c) -0.5 V. (d) Field H needed to assist domain wall nucleation (assisting field) plotted against the current that flows along line B during the pulse application.

Fig. 2. (a-c) Representative scope waveforms for different fields H and microwave frequencies. The microwave power is set to 0 dBm. (d) Scope signal amplitude plotted as a function of field H for various $f_{RF}$. The arrow illustrates the "assisting field" for $f_{RF}$=3.0 GHz. (e) Assisting field versus the current amplitude that flows through line A when the microwave current is applied. The arrow shows the point that corresponds to the arrow shown in (d).

Fig. 3. (a) The nucleation field, i.e. the sum of the Oersted field and the assisting field (shown in Figs. 1(d) and 2(e)), plotted as a function of the Oersted field at the field generation line. Squares: parallel pulse field, circles: transverse pulse field, up triangles: transverse microwave field (0.1 GHz), down triangles: transverse microwave field (3.0 GHz). (b) Nucleation field versus the microwave frequency when the microwave Oersted field at line A is set to ~27 Oe (microwave power of 0 dBm), as shown by the vertical dotted line in (a). (c) Optical micrograph of the device used to study the ferromagnetic resonance frequency. The blue dotted lines represent the permalloy Hall bar. (d) Microwave frequency applied to the waveguide versus the absolute mean of the field where the largest peaks are observed in the FMR measurements. Line is guide to the eye. Inset: wire resistance versus magnetic field when microwave current (frequency: 9.5 GHz) is applied to the waveguide.

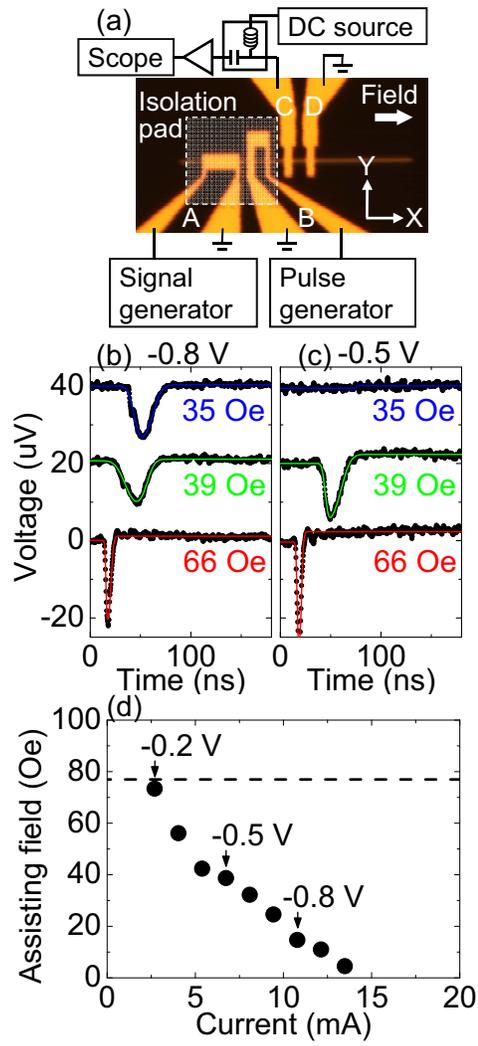

Fig. 1

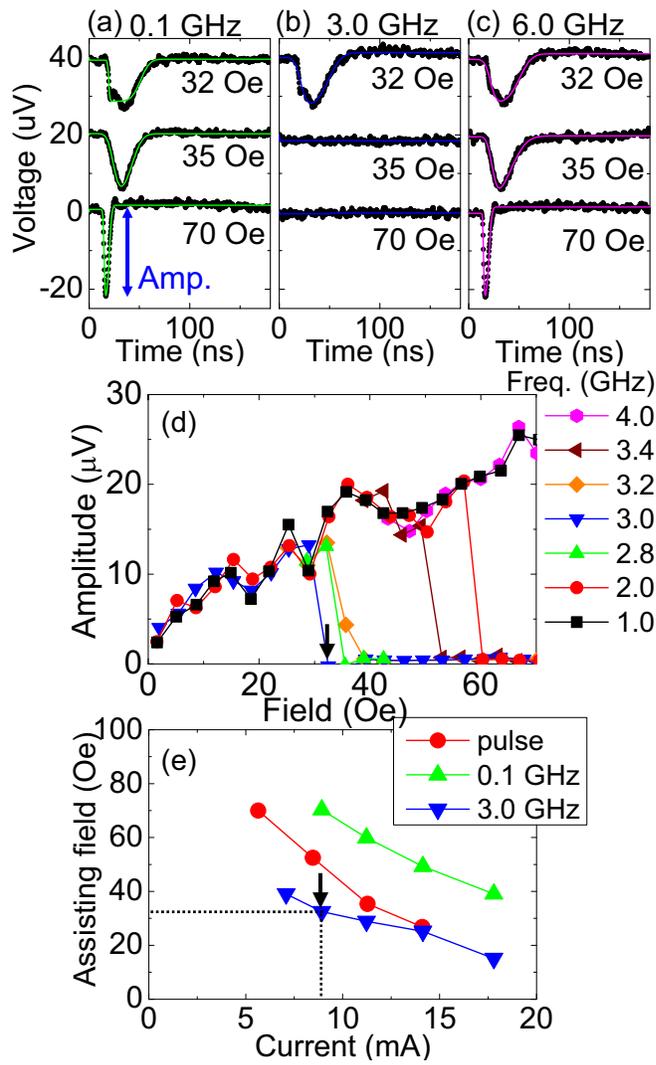

Fig. 2

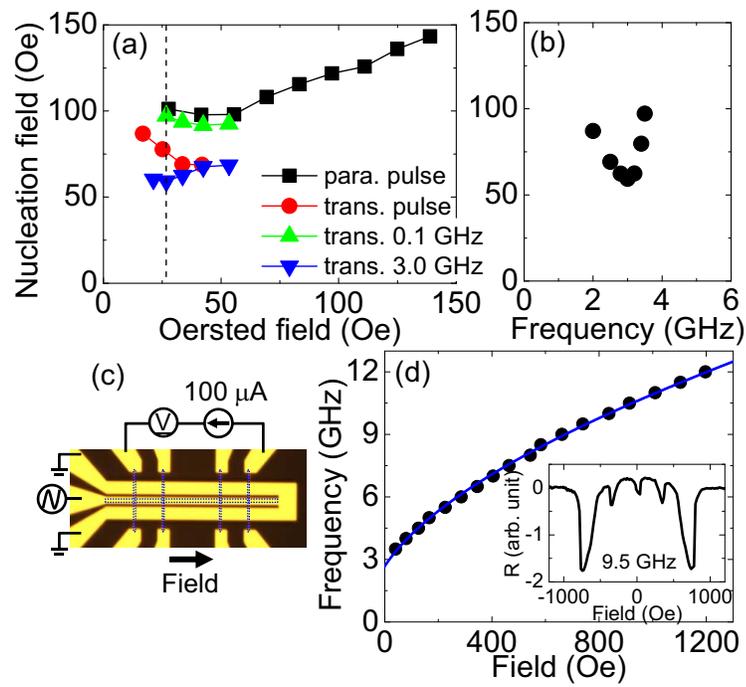

Fig. 3